\documentclass[]{spie}  %>>> use for US letter paper
\usepackage{etoolbox}
\csundef{ver@cite.sty}
\usepackage{amsmath,amsfonts,amssymb}
\usepackage{graphicx}
\usepackage[colorlinks=true, allcolors=blue]{hyperref}
\usepackage{gensymb}
\usepackage{wrapfig}
\usepackage{orcidlink}
\usepackage{subcaption}
\usepackage[backend=bibtex,style=numeric,sorting=none,natbib=false]{biblatex}
\title{CuRIOS-ED: The Technology Demonstrator for the CubeSats for Rapid Infrared and Optical Surveys Mission}

\author[a,b, *, \orcidlink{0000-0002-0699-2544}]{Hannah C. Gulick}
\author[a, \orcidlink{0000-0001-9611-0009}]{Jessica R. Lu}
\author[c]{Aryan Sood}
\author[a,b]{Steven V. W. Beckwith}
\author[a]{Charles-Antoine Claveau}
\author[a]{Joshua S. Bloom}
\author[b]{Kodi Rider}
\author[a, b]{Dan Werthimer}
\author[a, b]{Wei Liu}
\author[a]{Guy Nir}
\author[d]{Harrison Lee}
\author[b]{Jeremy McCauley}

\affil[a]{University of California, Berkeley, Department of Astronomy, Berkeley, CA, United States, 94720, USA}
\affil[b]{Space Sciences Laboratory, University of California, Berkeley, CA 94720, USA}
\affil[c]{Stanford University, Department of Mechanical Engineering, Stanford, CA, United States, 94305, USA}
\affil[d]{University of California, Berkeley, Department of Mechanical Engineering, Berkeley, CA, United States, 94720, USA}

\authorinfo{Send correspondence to H.C.G or J.R.L.: \\H.C.G. E-mail: hannah\_gulick@berkley.edu \\ J.R.L. E-mail: jlu.astro@berkeley.edu}

\begin{document} 
\maketitle

\begin{abstract}
The rise of time-domain astronomy including electromagnetic counterparts to gravitational waves, gravitational microlensing, explosive phenomena, and even astrometry with Gaia, are showing the power and need for surveys with high-cadence, large area, and long time baselines to study the transient universe. A constellation of SmallSats or CubeSats providing wide, instantaneous sky coverage down to 21 Vega mag at optical wavelengths would be ideal for addressing this need. We are assembling CuRIOS-ED (CubeSats for Rapid Infrared and Optical Survey--Exploration Demo), an optical telescope payload which will act as a technology demonstrator for a larger constellation of several hundred 16U CubeSats known as CuRIOS. The full CuRIOS constellation will study the death and afterlife of stars by providing all-sky, all-the-time observations to a depth of 21 Vega magnitudes in the optical bandpass. In preparation for CuRIOS, CuRIOS-ED will launch in late 2025 as part of the 12U Starspec InspireSat MVP payload funded through the Canadian Space Agency. CuRIOS-ED will be used to demonstrate the $<$1" pointing capabilities of the StarSpec ADCS system and to space-qualify a commercial camera package for use on the full CuRIOS payload. The CuRIOS-ED camera system will utilize a Sony IMX455 CMOS detector delivered in an off-the-shelf Atik apx60 package which has no previous space heritage. We deconstructed and repackaged the apx60 camera to make it compatible with operations in vacuum environments as well as the CubeSat form factor, power, and thermal constraints. By qualifying this commercial camera solution, the cost of each CuRIOS satellite will be greatly decreased ($\sim100\times$) when compared with current space-qualified cameras with IMX455 detectors. Therefore, the results from this work have great implications on the CuRIOS mission as well as other Cube or SmallSat missions. We discuss the CuRIOS-ED mission design with an emphasis on the disassembly, repackaging, and testing of the Atik apx60 for space-based missions. The testing results include characterization of the Sony IMX455 detector and Atik electronics performance. We find a read noise of 2.43$\pm$0.05 e- at a gain of 1 electron/ADU and detector temperatures ranging from -10 C to 25 C. The apx60's dark current is well below an electron per second at the temperatures and exposure times tested. The apx60 camera also exhibits patterned noise in the form of horizontal striping and an asymmetric signal gradient which increases across the detector's columns. We will also comment on preliminary environmental testing results.

\end{abstract}


\section{INTRODUCTION}
\label{sec:intro}  % \label{} allows reference to this section

A space-based survey with good sky-coverage, high cadence, and precise and stable photometry is needed to observe transient phenomena associated with the birth and evolution of black holes (BHs), neutron stars (NS), and white dwarves (WDs). We have developed a concept for a constellation of CubeSats called \textbf{Cu}beSats for \textbf{R}apid \textbf{I}nfrared and \textbf{O}ptical \textbf{S}urveys (CuRIOS) to address this need. CuRIOS uses an array of several hundred CubeSats with wide-field optical telescopes to provide wide-field, instantaneous sky coverage and study time-domain events. Figure \ref{fig:param_space} shows CuRIOS's expected performance compared to other novel observatories. CuRIOS will achieve a greater depth than current space-based surveys while also maintaining a larger FOV than leading ground based observatories. Additionally, as a space-based mission with a small solar keep-out-angle, the CuRIOS mission will have access to 270$\degree$ on the sky (75\% of the sky) at any given time. The bottom panel of Figure \ref{fig:param_space} highlights the large range of science cases covered by CuRIOS's broad cadence. The CuRIOS mission is described further in Section \ref{sec:CuRIOS_overview}. %A full mission outline is included in \cite{Gulick2024}.

The CuRIOS concept faces two main technological challenges. The first challenge is the lack of availability of a low-cost, space-qualified optical camera system. The rapid development and mass production of detector technology---such as large format ($>$100 MP), low noise complimentary metal-oxide-semiconductor (CMOS) detectors---has greatly expanded the potential of astronomical surveys by enabling instantaneous observations of fast and dim transients across large fractions of the sky. However, the development and testing of camera readout electronics suitable for space-based usage rapidly increases the cost, resulting in an off-the-shelf, space-grade camera package exceeding \$500k \footnote{As determined through several different vendor quotes for space-grade cameras.}. This price-point poses a major challenge to building scalable, low-cost satellites. The second technological challenge is the availability of an attitude determination and control system (ADCS) capable of $<$1" pointing stability. Current commercial attitude control systems for small satellites are not capable of achieving stabilization to better than a few seconds of arc \cite{Pong2018, Knapp:2020}, but $<$ 1'' stability is required by CuRIOS to monitor variability in crowded regions such as the Galactic Center and to resolve transients from their host galaxies. These systems are primarily limited by the accuracy of the star tracker's pointing measurements, thus a star tracker camera with finer pixels and higher accuracy could, in principle, greatly improve the pointing stability.

To address both of these challenges, we will launch a CuRIOS technology demonstrator called CuRIOS-Exploration Demo (CuRIOS-ED). CuRIOS-ED was selected to launch in 2026 as one of two payloads on a 12U CubeSat funded by the Canadian Space Agency (CSA). The CubeSat was awarded to StarSpec\footnote{A Canadian company specializing in ADCS systems (https://www.starspectechnologies.com)} as a technology demonstrator for a cutting-edge ADCS system capable of sub-arcsecond pointing stability. CuRIOS-ED will function to space-qualify a commercial camera system utilizing a Sony IMX455 detector. The camera package---an Atik apx60---was purchased as an off-the-shelf unit and does not have any space heritage. The camera was redesigned at the University of California, Berkeley (UCB) to comply with outgassing, volume, and mass constraints imposed by a space-based CubeSat. In addition to camera space-qualification, CuRIOS-ED will validate the StarSpec ADCS system using the high-accuracy IMX455 detector.

The full CuRIOS mission is outlined further in Section \ref{sec:CuRIOS_overview}, while the CuRIOS-ED instrument design and expected performance values are highlighted in Section \ref{subsec:CuRIOSED} and Section \ref{subsec:performance}, respectively. Details on the camera redesign as well as detector characterization and environmental testing are included in Section \ref{sec:camera}. The work highlighted in this paper is summarized in Section \ref{sec:conclusions}.

\section{Overview of the CuRIOS Concept}\label{sec:CuRIOS_overview}

\begin{figure} [ht]
   \begin{center}
   \begin{tabular}{c} %% tabular useful for creating an array of images 
   \includegraphics[height=10cm]{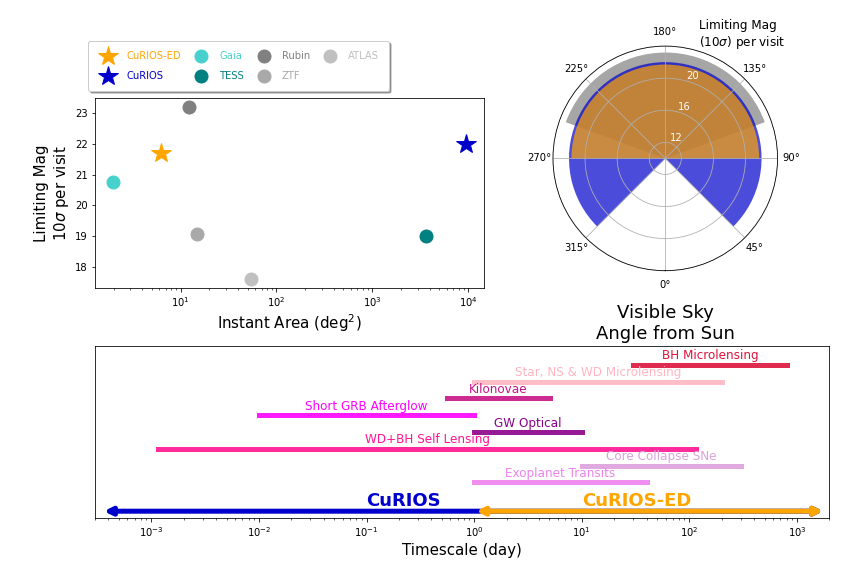}
   \end{tabular}
   \end{center}
   \caption[example] 
%>>>> use \label inside caption to get Fig. number with \ref{}
   { \label{fig:param_space} 
The CuRIOS and CuRIOS-ED Parameter Spaces. [upper left] The survey with the full CuRIOS constellation has a larger on-sky area than ground-based surveys and greater depth per visit than current space-based surveys. CuRIOS-ED has a smaller on-sky area comparable to leading ground-based surveys, while still achieving a similar depth to CuRIOS. [upper right] CuRIOS (blue region) provides much higher instantaneous sky coverage due to its shark-nose baffle which allows each satellite to observe within 45$\degree$ of the Sun. Without this baffle, CuRIOS-ED (orange region) will be limited to observing within 90$\degree$ of the Sun, however this still provides a higher instantaneous sky coverage than ground-based telescopes like Rubin (grey region). The circular contours highlight the 10$\sigma$ limiting magnitude for a single visit. While slightly shallower than Rubin (limiting Vega magnitude = 23), CuRIOS and CuRIOS-ED still achieve a competitive limiting Vega magnitude of 21.7 and 21.6, as indicated by the radius of the blue and orange areas, respectively. Coupled with its larger instantaneous sky coverage, CuRIOS's depth, cadence, and duration make it a powerful tool for detecting a large range of science cases [bottom] while CuRIOS-ED is more equipped for slower cadence follow-up of longer timescale events---such as stellar mass black hole microlensing. The timescales of a selection of transient phenomena are compared to the CuRIOS-ED (orange arrow) and CuRIOS (blue arrow) cadences and total mission durations.}
\end{figure}

The concept for CuRIOS consists of a constellation of $\sim$300, 16U CubeSats that provides simultaneous and continuous observations of a large fraction of the sky at $\sim$15 minute cadence. Each CubeSat will have an $\sim$35--40 deg$^2$ FOV with $\sim$1--2.6” resolution across the field. CuRIOS will observe sources down to R = 21st Vega magnitude in 15 minute stacked exposures with an SNR = 10, as summarized in the second column of Table \ref{tab:characteristics}. Figure \ref{fig:param_space} shows CuRIOS's expected performance compared to other wide-area observatories. CuRIOS will achieve a greater depth than current space-based surveys while also maintaining a larger FOV than leading ground based observatories. Additionally, as a space-based mission with a small solar keep-out-angle, the CuRIOS mission will have access to 270$\degree$ on the sky at any given time. The bottom panel of Figure \ref{fig:param_space} highlights the large range of science cases covered by CuRIOS's broad cadence. %\cite{Gulick2024} contains a full description of the mission design.

\begin{table}[t!]
\caption{CuRIOS$^a$ \& CuRIOS-ED Instrument Specifications} 
\label{tab:characteristics}
\begin{center}       
\begin{tabular}{l|l|l}
\hline
\rule[-1ex]{0pt}{3.5ex}  Property & CuRIOS & CuRIOS-ED  \\
\hline\hline
Bandpass & 450 nm to 800 nm & 450 nm to 800 nm  \\
Aperture & 15 cm & 9.5 cm \\
Resolution (core FWHM) & 1.5'' & 2.06'' \\
 Plate Scale & 2.6'' / pix  & 1.33'' / pix \\
             & 3.7 $\mu$m / pix &  1.16" / pix\\
FoV & 6.9$\degree$ $\times$ 4.6$\degree$ & 2.5$\degree$ $\times$ 2.5$\degree$ \\
Sky Coverage &  32 deg$^2$ & 6.25 deg$^2$\\
Limiting  Magnitude & R-band = 21.7 mag & CuRIOS-ED Band$^a$ = 21.6 mag  \\
\textit{(Vega, 10$\sigma$ in 900 s)} & & \\
Photometric Precision & 0.1 mag at R = 21 & 0.1 mag at CuRIOS Band = 21  \\
\hline 
\end{tabular}
\end{center}
\end{table}

With CuRIOS, we address the 2020 Decadal Survey for Astronomy (Astro2020) call for a space-based, time-domain program capable of studying the dynamic Universe \cite{astro2020} by building a demographic census of stellar remnants in the Milky Way in isolated and binary systems. Additionally, we will observe the explosive creation or merger of stellar remnants in other galaxies. We also note that CuRIOS is highly complimentary to Roman \cite{Johnson2020} and Rubin \cite{Hambleton2023}---two other time domain surveys that will run concurrently---as CuRIOS has a bigger field of view and higher sky coverage that allows it to fill the temporal gaps between Roman and Rubin observations. The CuRIOS survey will be shallower (limiting magnitude of R = 21) than either Roman or Rubin, and will therefore only share observations of the brightest transients. While the full CuRIOS constellation design is still in progress, an initial design for the first three CuRIOS CubeSats---referred to as CuRIOS--Initial Demo (CuRIOS-ID)---is nearly complete.% \cite{Gulick2024} and will be submitted to an upcoming proposals.

\section{The CuRIOS Exploration Demo: CuRIOS-ED}\label{sec:CuRIOS}

The CubeSats for Rapid Infrared and Optical Surveys–Exploration Demonstrator (CuRIOS-ED) is a scaled-down version of the CuRIOS payload that will serve two primary goals: validating the StarSpec ADCS pointing system and space qualifying a low-cost, commercial camera system. Both of these demonstrations will be fundamental for the success of the future CuRIOS mission. A secondary goal will be to provide wide-field, high-resolution images to assist in  calibration and pipeline development for the full CuRIOS mission. These images will also be used to validate the StarSpec ADCS pointing stability and to study microlensing from stellar mass black holes in the Galactic Center.

CuRIOS-ED will fly on the StarSpec InspireSAT MVP 12U CubeSat, orbiting in low-earth orbit (LEO) along a sun-synchronous (SSO) trajectory. The payload will launch in 2026 on a SpaceX RideShare. The bus, radio, bus computer, and power supplies will be procured and assembled by StarSpec. StarSpec will also build the second payload which will demonstrate an ADCS system capable of $<$1" pointing stability in LEO, as required by CuRIOS and many upcoming small satellite missions. The CuRIOS-ED payload will be provided by the University of California, Berkeley (UCB) and utilizes mainly off-the-shelf parts.

\subsection{Instrument Design}\label{subsec:CuRIOSED}

\begin{figure} [ht]
   \begin{center}
   \begin{tabular}{c} %% tabular useful for creating an array of images 
   \includegraphics[height=8cm]{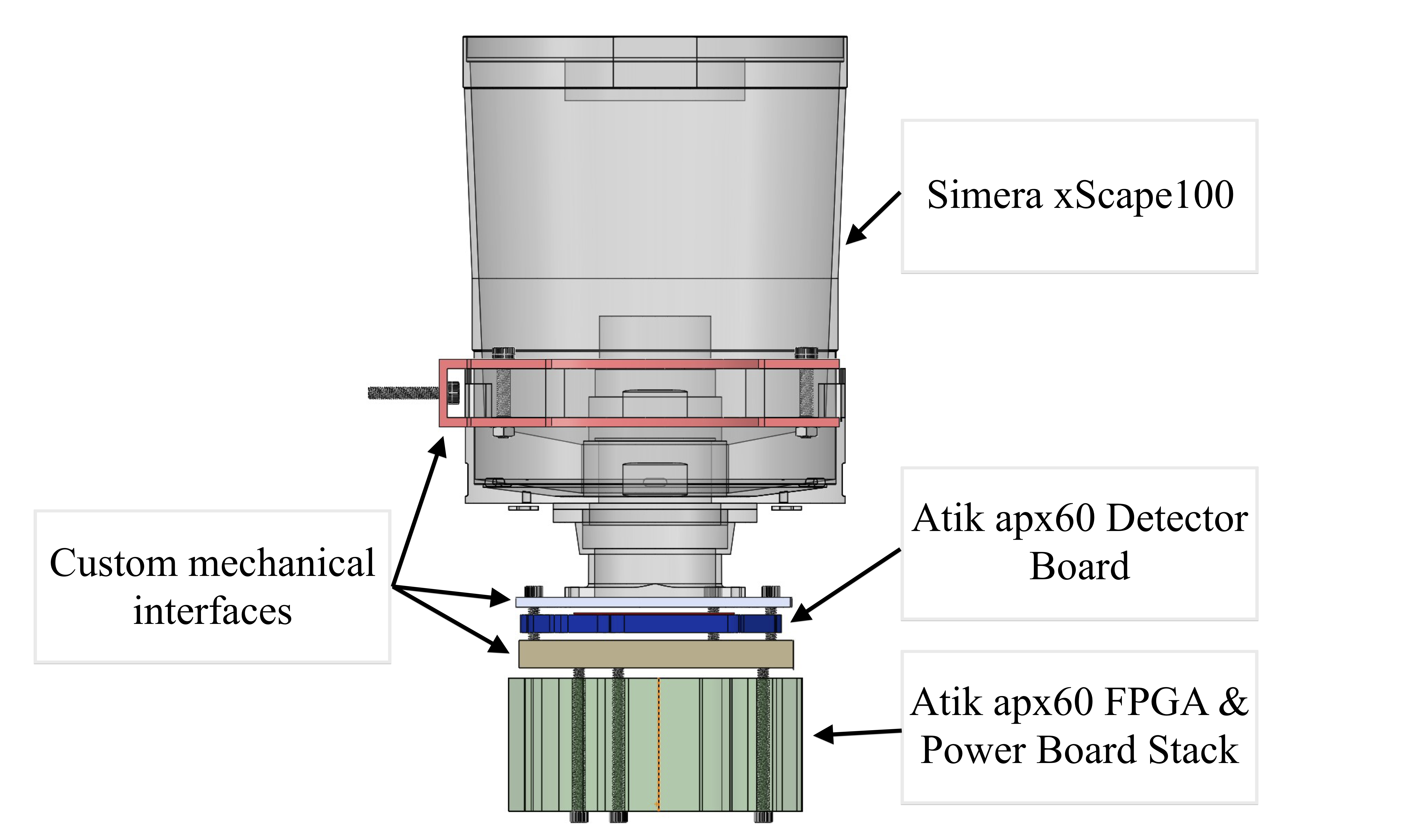}
   \end{tabular}
   \end{center}
   \caption 
   { \label{fig:CAD} 
The CuRIOS-ED payload including a Simera xScape100 optical front-end and repackaged Atik apx60 camera with Sony IMX455 CMOS detector. The custom mechanical interfaces were designed at UCB and function to mount the camera electronics to the Simera optic as well as the optic to the bus structure. The entire payload fits within a 10 cm $\times$ 10 cm $\times$ 20 cm volume.}
\end{figure}

The CuRIOS-ED payload was allocated a 1Ux2U (10 cm $\times$ 10 cm $\times$ 20 cm) volume on the InspireSat 12U bus and will include a telescope, payload computer (CPU), and optical CMOS camera. Figure \ref{fig:CAD} shows the payload CAD with components labeled. CuRIOS-ED will not include any filters, and will instead collect information across the full 450 nm to 800 nm bandpass. We refer to this as the 'CuRIOS filter' in Table \ref{tab:characteristics}.

%The CuRIOS-ED payload was allocated a 1Ux2U (10 cm $\times$ 10 cm $\times$ 20 cm) volume and will include a 95 mm Simera xScape100 Optical Front-End (OFE) with a Schmidt-Cassegrain telescope design, an optical CMOS camera system, and an EnduroSat Onboard Computer (OBC) as the payload computer. The CMOS camera will utilize a Sony IMX455 and the readout electronics from the off-the-shelf Atik apx60 camera. Figure \ref{fig:CAD} shows the science instrument CAD model with components labeled. CuRIOS-ED will not include any filters, and will instead collect information across the full 450 nm to 800 nm bandpass. We refer to this as the 'CuRIOS filter' or Cu for short (ref. Table \ref{tab:characteristics}).

The payload telescope will be a Simera xScape100 VNIR (visible and near-infrared) optical front-end (OFE) with a Schmidt-Cassegrain design. The xScape100 was purchased as a stand-alone product without the standard Simera camera package attached. The optic is optimized to fit within a 3U CubeSat package while maintaining a 95 mm optical aperture and comes with a bus-to-optic mechanical interface. The OFE is space-qualified---having been shown to survive the structural load inflicted during launch/deployment as well as the harsh conditions (such as radiation damage) experienced on-orbit. The xScape100's transmission throughput remains constant at $\sim$65\% across the full CuRIOS-ED band while maintaining a 2.5$\degree$ $\times$ 2.5$\degree$ and 2.06'' spatial resolution (see Table \ref{tab:characteristics}).

The first choice for payload CPU is the Xiphos 7S computer \footnote{\url{https://xiphos.com/product-details/q7}}. We are exploring whether a lower-cost Raspberry Pi could be a solution.

The CuRIOS-ED payload will utilize a repackaged version of the commercial Atik apx60\footnote{\url{https://www.atik-cameras.com/large-format-cooled-cmos/apx-60/}} camera with IMX455 CMOS detector---the same detector intended for use on the full CuRIOS mission. Characteristics for the apx60 are included in Table \ref{tab:characteristics}. To fit within the CubeSat form factor and minimize power consumption, additional tech (i.e. thermal controls) as well as the apx60's external, metal casing was removed from the camera. The final CuRIOS-ED camera consists of the readout and power electronics and fits within an 8 cm $\times$ 8 cm $\times$ 5 cm volume.

Finally, we designed the CuRIOS-ED payload to be a proof-of-concept for the full CuRIOS payload, and therefore kept the payloads as similar as possible. We note that the telescope will be the main difference between the CuRIOS-ED and CuRIOS payloads. While both telescopes utilize a Schmidt-Cassegrain telescope design, CuRIOS-ED's telescope is an off-the-shelf, classical OFE and CuRIOS's telescope is a novel monolithic optic. % discussed further in \cite{Gulick2024}.

\subsection{Operational Modes}\label{subsec:modes}

On-orbit, CuRIOS-ED will operate in two different operational data modes: (1) science and (2) ADCS validation. The \textbf{science mode} will utilize the full-frame array with a 30-second exposure time to capture science images of sources such as the Orion Nebula, globular clusters, and the Galactic Plane. These images will be used to provide visual proof of the StarSpec ADCS system by showing astrophysical objects with and without the improved ADCS corrections. These images will also be used to communicate the mission to the public and to study BH microlensing rates in the Milky Way.

To reduce the data volume being produced by CuRIOS-ED---and thus the required downlinking bandwidth---the science mode images will be stacked on-board the spacecraft. Nominally, the individual 30 second images will be stacked to a total integration time of 15 minutes. However, the depth of the stack (e.g. 10 minutes instead of 15 minutes) can be modified on-orbit to enable the study of science cases requiring a finer time resolution. Only the stacked image will be sent to ground, thus reducing the data volume by a factor of 30 for a total stacked time of 15 minutes.

The \textbf{ADCS validation mode} will use a smaller subarray of 16 x 16 pixels or multiple sub-arrays and a readout rate of 30 Hz to provide high-speed pointing information on the StarSpec ADCS system. In the validation mode, CuRIOS-ED will measure the RA and Dec of bright stars within the subarray and send these coordinates to the StarSpec bus to be utilized by the ADCS system. The validation mode has two varieties: (1) validation-cutout and (2) validation-centroid. In the validation-cutout mode, the 16 x 16 pixel cutout images are saved and sent in raw format to the ground. In the validation-centroid mode, the 16 x 16 pixel cutout images are processed and the flux-weighted centroids are extracted. In this mode, only the time of the observation and centroid (x, y) coordinates are sent to the ground.

\subsection{Instrument Performance}\label{subsec:performance}

We calculate the signal-to-noise (SNR) ratio vs.~brightness for the CuRIOS-ED science and validation modes with 30 second and 0.03 second exposure times, respectively. We find the SNR for CuRIOS-ED taking into account sky brightness, telescope throughput, detector quantum efficiency, and spectral radiance following the equation in \cite[see][Eq.2]{Zhao2021}, with an additional component for emission from background sources such as that from a host galaxy or star field.

%We use the SNR code outlines in Gulick et al. in  prep which finds the sensitivity for optical/infrared telescopes taking into account sky brightness, telescope throughput, detector quantum efficiency, and spectral radiance following the equation in \cite[see][Eq.2]{Zhao2021}, with an additional component for emission from background sources such as that from a host galaxy or star field.

For each flux source (e.g. target or background), we use CuRIOS-ED's circular telescope aperture with a diameter of 9.5 cm and a photometric extraction box size of 4'' $\times$ 4''. Obscuration from the telescope's secondary mirror is removed from the total usable aperture by subtracting the obscured area from the total aperture area, leaving a usable area of 53.4 cm$^2$. The signal is calculated over the entire CuRIOS-ED bandpass of 450 nm to 800 nm, where the instrumental response across the bandpass is set to 65\% to account for the xScape100's trasmittance (ref. Section \ref{subsec:CuRIOSED}) and further scaled by the quantum efficiency (QE) of the detector as measured in \cite{Gill:2022}. The central wavelength and width of the filter are calculated to be 600 nm and 350 nm, respectively. The signal is converted to electrons via Eq. 3 in \cite{Zhao2021}. The sky background, m\textit{$B_{sky}$}, was calculated using the combined Earth-shine and zodiacal background rates given in \cite{Dressel22}.  $B_{ins}$ was given by the spectral radiance assuming blackbody emission and a temperature of 273.15 K. \textit{$B_{sky}$} and $B_{ins}$ were converted to a signal via Eq. 4 \& 5 in \cite{Zhao2021}.

%We use the optic's instrumental throughput of 65\% as the total instrumental throughput and a detector QE of $\eta = 0.65$ as measured in \cite{Gill:2022}. Note that both the instrumental throughput and QE were used to calculate the CuRIOS filter response profile, therefore   

\begin{figure} [ht]
   \begin{center}
   \begin{tabular}{c} %% tabular useful for creating an array of images 
   \includegraphics[height=8cm]{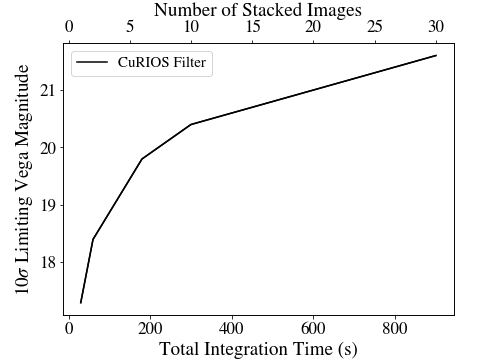}
   \end{tabular}
   \end{center}
   \caption 
   { \label{fig:SNR_sci} 
The limiting Vega magnitude for a 10$\sigma$ detection versus the total detector integration time. The total integration time is composed of stacked 30 second exposures---therefore a 900 second total integration time corresponds to and image with 30 stacked exposures. The number of stacked images required to reach each total integration time is shown on the top x-axis.}
\end{figure}

Figure \ref{fig:SNR_sci} shows the limiting magnitude for a 10$\sigma$ detection as a function of the total, stacked integration time when operating in science mode. The total integration time is composed of stacked, individual 30-second exposures. An additional x-axis is included on top of the figure to show the number of 30-second exposures being stacked at each total integration time. At a total integration time of 30 seconds (i.e. one exposure, unstacked), CuRIOS-ED reaches a limiting magnitude of 17.3 Vega mags with a 10$\sigma$ detection. At a total integration time of 900 seconds (i.e. a stack of 30 exposures), CuRIOS-ED reaches a limiting magnitude of 21.6 Vega mags.

While the CuRIOS telescope aperture (15 cm) is larger than the CuRIOS-ED aperture (9.5 cm), we find that CuRIOS-ED's limiting magnitude is comparable to CuRIOS's limiting magnitude in the R-band (see Table \ref{tab:characteristics})). This is due to a difference in CuRIOS and CuRIOS-ED's observable bandpasses and percentage of telescope obscuration. CuRIOS will utilize an R-band filter while CuRIOS-ED will collect light over the full 450 nm to 800 nm, thus restricting the bandpass over which CuRIOS collects signal photons. Additionally, the CuRIOS monolithic optic suffers from substantial obscuration caused by its secondary mirror. Due to these two factors, the SNR profiles between CuRIOS and CuRIOS-ED are comparable despite their different aperture sizes.

Additionally, they are significantly deeper than current space-based surveys such as the Transiting Exoplanet Survey Satellite \cite[TESS;][]{Ricker2015} with a 3$\sigma$ limiting magnitude of 19 in a 10 minute exposure and Kepler \cite{Borucki2007} with a larger aperture ($\sim$1 m) a limiting Kepler magnitude of 19. The increased depth in the CuRIOS satellites when compared to Kepler especially, is due to survey design choices including FOV size and detector selections. Kepler's primary purpose is to precisely measure the brightness of stars over long periods to detect exoplanet transits. With an extremely large FOV of 115$\degree$, Kepler's pixel scale of 3.98"/pixel is relatively large. Therefore, light from a point source is diluted over the large area covered by a single pixel, thus reducing the sensitivity to dim objects. Secondly, Kepler has a relatively high readout noise of 100 e- \cite{Gilliland2011} which is balanced by the telescope's long integration times but still places an instrumental limit that restricts deep observing.

\section{Camera}\label{sec:camera}

Through CuRIOS-ED, we intend to space-qualify the repackaged apx60 camera (see Section \ref{subsec:CuRIOSED}) for use on the future CuRIOS CubeSats. The price of existing space-qualified camera packages is prohibitive to building scalable, low-cost satellites. For a constellation of satellites, this cost rapidly compounds so the use of a low-cost, commercial solution is vital. By utilizing the repackaged apx60 camera, we reduce the cost per camera unit by a factor of $\sim$100 when compared to space-qualified cameras utilizing the same IMX455 detector. However, the apx60 is designed for ground-based, amateur astronomy applications and has no previous flight heritage or space-qualification testing. Any number of issues can arise when sending a commercial package to space, including outgassing from glues, electrical component failure from temperature or radiation extremes, electrical point-of-contacts disconnecting from vibrations, or overheating of components. Therefore, intensive characteristic and environmental testing is needed to prepare a commercial camera for use in space. Our environmental testing results are discussed further in Section \ref{subsec:environ_test}.

\subsection{Performance Characterization of Atik apx60}

We performed comprehensive noise characterization using the off-the-shelf apx60 package before disassembly. These values are taken to be the detector's nominal performance and all subsequent environmental testing performance results are compared to this baseline. The characterizations discussed in this section include: read noise, dark current, and patterned noise.

\subsubsection{Baseline Image Analysis}\label{subsec:intensity}

Dark frame images were taken with the apx60 at a range of detector temperatures and exposure times as part of the baseline camera characterization. The detector temperature was controlled with the internal apx60 fan and Peltier cooler, with setpoints of -10 C, -5 C, 0 C, 5 C, 10 C, and 15 C. Additional data were taken at 'ambient' temperatures with the cooler off which corresponds to 19 C. At each detector temperature, dark frames with exposure times of 0.001 s, 0.01 s, 0.1 s, 1 s, 5 s, 10 s, 15 s, and 30 s were taken. All dark frames used in this section were taken with a gain of 1 and an offset of 200 ADU---where the detector offset functions to prevent negative pixel values and 200 ADU is the nominal offset value set by Atik. The dark frames were taken in a dark room with the included apx60 cap screwed on to the camera housing and blocking any light from reaching the detector.

%We analyzed the dark frames by calculating the mean pixel value across the images' columns and rows.

\begin{figure} [ht]
   \begin{center}
   \begin{tabular}{c} %% tabular useful for creating an array of images 
   \includegraphics[height=7cm]{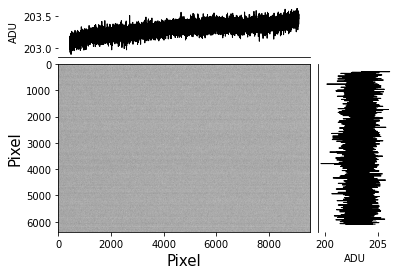}
   \end{tabular}
   \end{center}
   \caption 
   { \label{fig:apx60_intensity} 
An example image from the Atik apx60. The image was taken with a 0.001 second exposure, gain of 1, offset of 200 ADU, and a detector temperature of 19 C (i.e. ambient with no active temperature control). The top and right-most panels show the mean pixel signal across the columns and rows, respectively. }
\end{figure}

We examined the dark frames for amplifier glow or other types of spatially structured signals by calculating the mean pixel value across the images' columns and rows.
%We characterize the baseline intensity profile in the dark frames by calculating the mean pixel value along each row and column.
Figure \ref{fig:apx60_intensity} shows an example apx60 image with an exposure time of 0.001 seconds, gain of 1, offset of 200 ADU, and temperature of 19 C (i.e. ambient with no active cooling or heating). The panels on top and to the right of the central figure show the mean column and row pixel intensities in ADU, respectively. Notably, this figure reveals the apx60 has a small but measurable asymmetric signal profile across the columns, which increases with column number. We fit the signal profile in the top panel of Figure \ref{fig:apx60_intensity} with a linear regression to get the slope and intercept of the asymmetry. This fit is performed for five images within a given temperature or exposure time dataset, and the errorbars are given as the STD of the fit values across the five images. Performing this analysis on 0.001 s dark frames with different detector temperatures, we find the y-intercept of the asymmetry (or minimum signal) increases with temperature from 201.95$\pm$0.013 ADU at -10 C to 203.17$\pm$0.010 ADU at 19 C. As a function of temperature, the slope of the signal asymmetry stays consistent within the errorbars at $\sim$3.3e-5$\pm$3e-6 ADU/s, corresponding to an $\sim$0.3 ADU signal magnitude. Similarly, when analyzing data with a detector temperature of 19 C and exposure times varying from 0.001 s to 30 s, the y-intercept of the asymmetry increases with exposure time from 203.17$\pm$0.010 ADU at 0.001 s exposure to 204.21$\pm$0.017 ADU at 30 s exposures. However, we find that the signal asymmetry's slope exhibits a slight exposure dependency, increasing from 2.89e-5$\pm$3e-7 ADU/s at 0.001 s exposures to 4.29e-5$\pm$6e-7 ADU/s at 30 s exposures. This corresponds to a 0.133 ADU change in the magnitude of the signal's asymmetry from 0.001 s to 30 s and suggests there is a weak time-dependent signal present in the dark frames.

%no variation means it is a source of electronics noise or amplifier glow, even though every pixel has own amplifier they are offloaded to second stage on one side--need to make a conclusion
%if not flat then ambient light coming

The right-most panel of Figure \ref{fig:apx60_intensity} shows the mean signal across the image's rows. As is discussed further in Section \ref{subsec:patnoise}, the row signal is dominated by patterned noise which appears as horizontal stripes which move from frame to frame. The magnitude of the patterned noise is much larger than that of the signal asymmetry across the columns, with a min to max magnitude of 6$\pm$0.4 ADU for exposure times of 0.001 s and detector temperature of -10 C. This value increases by $\sim$1 ADU as exposure times are increased to 30 s or detector temperatures are increased to 19 C. Figure \ref{fig:pattern_noise} visually highlights the patterned noise present.

Furthermore, we find the dark frame data have a median value of $\sim$203 ADU (no outlier cuts) for an exposure time of 0.001 seconds at room temperature (19 C). Subtracting the 200 ADU offset and converting to electrons, this corresponds to an electronics noise floor of 3 electrons at 19 C. We note that this floor decreases to $\sim$2 electrons at -10 C. A total of $<$0.0001\% of pixels have outlying values $>$20$\sigma$ above or below this median value. These outlying pixels tend to vary across each frame and from frame-to-frame, suggesting most are random events such as cosmic rays. In a very few cases, a pixel is bad and remains well above or below the median value across all frames.

% QQQ The outlying pixels are also evenly distributed across the frames and not crowded together on one side of the image, confirming the absence of light leaks in the data or asymmetric signals in the data. Therefore, we determine the dark frames are adequate for read noise measurements. 
%All frames used to calculate read noise were taken in a completely dark room with the apx60 included cap fully covering the detector. The shortest possible exposure time (0.001 seconds) was used to remove contributions from dark current. 

\subsubsection{Read Noise}\label{subsec:readnoise}

Detector read noise arises from the electronics processes required to amplify and convert the detector's photoelectron signal into a readable voltage value. Read noise is inherent to any detector---i.e. CCD or CMOS---and typically sets the instrumental noise floor or sensitivity threshold in short exposures. While independent of exposure time and thus signal level, read noise will vary with the detector temperature and readout speed, with higher readout rates increasing the amount of read noise. This is largely due to the degradation of  amplifier and ADC performance, increased crosstalk between pixels, and increased thermal noise at higher readout frequencies \cite{Wanqing2002, Greffe2022, Wu2022}. 

%Additionally, due to the extremely low noise properties of modern CMOS detectors, it is possible for the electronics producing read noise to have a small dependence on the detector temperature \cite{Wu2022}.

To measure intrinsic read noise, it must be isolated from other signals or noise sources. Therefore, read noise was measured in dark frames with short 0.001 second exposure times. The short exposure acts to remove contributions from shot noise and dark current.
The read noise is calculated by taking the standard deviation (STD) over a range of values outputted by a detector's amplifier and readout electronics circuit. In CMOS detectors, each pixel has its own amplifier and readout electronics package requiring the read noise to be measured for each individual pixel. This differs from traditional CCD detectors where a single set of readout electronics is used to readout the entire detector thus imposing the same amount of read noise on each pixel and allowing for the STD to be taken across the entire image.

\begin{figure} [ht]
   \begin{center}
   \begin{tabular}{c} %% tabular useful for creating an array of images 
   \includegraphics[height=7cm]{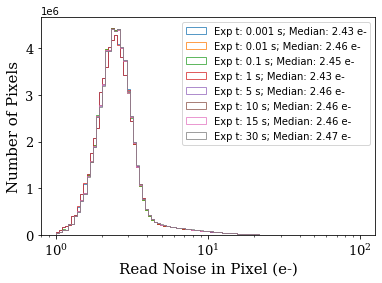}
   \end{tabular}
   \end{center}
   \caption 
   { \label{fig:readnoise_distrib} 
A distribution of the read noises in all 61 MP of the apx60 camera with Sony IMX455 detector for exposure times of 0.001, 0.01, 0.1, 1, 5, 10, 15, and 30 seconds. The detector temperature is set to 5 C and all frames included are within 0.2 C of the setpoint. The median read noise values for each exposure time are included in the legend. The error on each read noise value is $\pm$0.05 electrons. The distribution exhibits a long tail at high read noises caused by hot pixels. Therefore the x-axis utilizes logarithmic binning  to highlight the structure in the main distribution.  }
\end{figure}

In this analysis, we measure the read noise in the Sony IMX455 CMOS detector by calculating the read noise in each of the individual 61 mega-pixels. We begin by subtracting a second dark frame with the same detector temperature, gain and exposure time from the initial dark frame and divide the resulting image by the $\sqrt{2}$ to account for the noise added by the second dark frame. We then extract a single pixel's value from a series of 20 back-to-back, bias-corrected dark frames and calculate the STD of these values. We perform an outlier cut to remove any values outside of $\pm$6$\sigma$, though we do not find any pixels with outlying values in the dataset. Repeating this step for all 61 MP in the IMX455 creates a distribution of the individual pixel read noise values in the detector (see Figure \ref{fig:readnoise_distrib}). The distribution's median value is taken to be the detector's read noise.

%calculate the STD of a single pixel's value in a series of 10 back-to-back, bias-corrected 0.001 second dark frames.

\begin{figure}%
    \centering
    \subfloat{{\includegraphics[width=8cm]{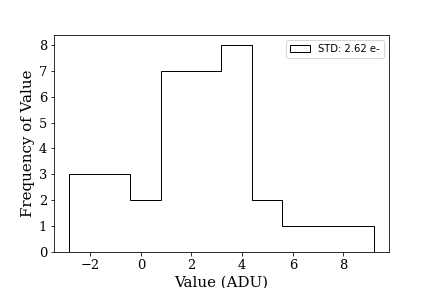} }}%
    \qquad
    \subfloat{{\includegraphics[width=8cm]{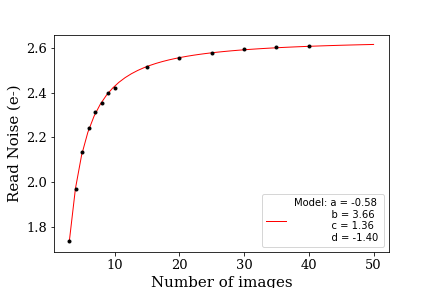} }}%
    \caption{[left] The distribution of values in a single pixel across 40 consecutive exposures. Each exposure has an exposure time of 0.001 seconds, a gain of 1, and was readout in fast mode. [right] The relationship between the median pixel read noise value and the number of images the pixel's read noise was calculated across. The trend is fit with a four point logistic function, with best-fit parameters included in the legend. 
    The standard error on the mean (SEM) is included for each read noise data point, however, the large number of pixels results in negligibly small errors.}%
    \label{fig:readnoise_plat}%
\end{figure}

However, in order to calculate the read noise for a single pixel across a series of frames, a minimum number of frames is needed to achieve good statistics. The left panel of Figure \ref{fig:readnoise_plat} shows a representative distribution of values in a single pixel across 40 frames. We find that a pixel's value distribution is often non-Gaussian, showing multiple peaks or asymmetry. Because of this, we find that in datasets with less than 20 frames, the measured read noise in a pixel increases as the number of frames in the analyzed dataset increases.

The right panel of Figure \ref{fig:readnoise_plat} shows the measured relationship between the number of images used to calculate an individual pixel's read noise and the corresponding median pixel read noise (i.e. detector read noise). The trend is calculated over datasets ranging from 3 images to 40 images. We fit this trend with a four part logistic model:

\begin{equation}
\mathrm{RN} = a + \frac{b}{1+(\frac{n_{img}}{c})^d} 
\end{equation}

\noindent where $n_{img}$ is the number of images in the analyzed dataset and RN is the corresponding median read noise value. The best-fit parameters are included in the figure's legend.

We find that datasets using less than 20 images underestimate the read noise value by $\sim$0.5 electrons for a dataset with 5 images and $\sim$0.2 electrons for a dataset with 10 images. This finding suggests using larger dataset with 50+ images results in very accurate read noise measurements, however, each IMX455 image is $\sim$122 MB and is expensive to process. Therefore, we adopt datasets with 20 images to provide a good balance between read noise accuracy and computational expense. With 20 images per dataset, the read noise is calculated within $\sim$0.05 electrons of the read noise plateau (see Figure \ref{fig:readnoise_plat}), therefore, we include this value as an additional error component in the following calculations.

Figure \ref{fig:readnoise_distrib} shows the individual pixel STD distributions (calculated over 20 images) for a range of exposure times at a detector temperature of 5 C. The exposure times include: 0.001 s, 0.01 s, 0.1 s, 1 s, 5 s, 10 s, 15 s and 30 s. The pixel read noise distributions agree within error for all exposure times tested. This means the read noise analysis method is independent of exposure time due to a low to negligible dark current. 

%In the figure, two distinct distribution shapes can be seen---a taller distribution associated with exposure times $<$ 1 second and a shorter, slightly wider distribution associated withe exposure times at 1 second and longer. This division likely is caused by the proprietary electronics which handle the image readout.

Additionally, we measure the read noise in the unmodified apx60 package as a function of detector temperature at a low gain setting with 1 electron/ADU. For these measurements, datasets with only 10 frames were available, therefore we assign a larger error of 0.2 electrons to the final results. The detector temperature is modified from -10 C to 15 C at intervals of 5 C using the Atik apx60 SDK which controls a proprietary cooling system in the apx60 package. The temperature control system is stable to 0.1 C, and all data included in a given temperature set are required to be within $\pm$0.2 C of the set value.

\begin{figure}%
    \centering
    \subfloat{{\includegraphics[width=7cm]{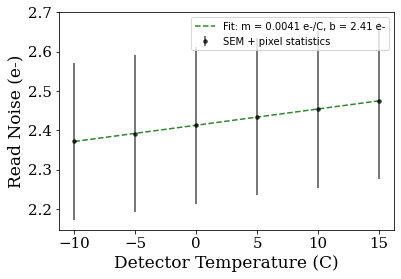} }}%
    \qquad
    \subfloat{{\includegraphics[width=8cm]{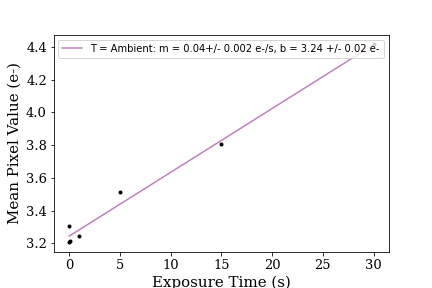} }}%
    \caption{[left] Read noise as a function of detector temperature at low gain = 1. The data are fit with a linear regression as shown by the green dashed line. The sum of the SEM errors and the 0.2 e- error from the data completeness calculation (see Section \ref{subsec:readnoise}) are included for each data point. Note that the SEM errors are very small due to the large number of pixels, therefore the constant 0.2 e- data completeness error dominates the error bars. The profile is best fit with a linear regression with a slope of 4.1e-3 e-/C and intercept of 2.41 e-. [right] The mean pixel value over an image versus the exposure time. The original image offset of 200 ADU was subtracted from the image to provide the values in units of electrons. The mean pixel value is calculated over a single image and then averaged over 10 frames with the same exposure time. The detector temperature is resting at ambient ($\sim$19 C). The SEM errors are included on each data point, however, due to the number of pixels in the detector these errors are very small. The profile is fit with a linear regression with a slope and intercept parameter of 4e-2 e-/s and 3.24 e-, respectively. Section \ref{subsec:DC} includes justification on the temperature set point. }%
    \label{fig:readnoise}%
\end{figure}

Figure \ref{fig:readnoise} shows the read noise as a function of detector temperature in the left panel. The read noise values are nicely fit with a linear regression profile as indicated by the green dashed line. At low gain settings, the read noise is 2.43$\pm$0.05 electrons across the temperature range tested. The read noise in the IMX455 does have a small dependence on temperature. However, the impact of this dependence on the total noise in the system is much less than an electron and therefore, negligible.

%\begin{figure}%
%    \centering
%    \subfloat[\centering Low Gain = 1 electron/ADU]{{\includegraphics[width=7cm]{Figures/CMOSanalysis_read_noise_v_det_temp.png} }}%
%    \qquad
%    \subfloat[\centering High Gain = 25 electron/ADU]{{\includegraphics[width=7.2cm]{Figures/CMOSanalysis_read_noise_v_det_temp_highgain.png} }}%
%    \caption{Read noise as a function of detector temperature at low and high gain settings. The data are fit with a linear regression as shown by the green dashed line. QQQ SEM errors are included for each data point, though the errors are very small and therefore not visible in these figures.}%
%    \label{fig:readnoise}%
%\end{figure}

\subsubsection{Dark Current}\label{subsec:DC}

Detector dark current arises from thermal fluctuations in a detector from electronics or the surrounding medium. As a detector exposes and readouts out images, the heat from the electronics causes thermal electrons to build-up on the sensor. These electrons are read out with the true signal. Dark current can be measured in dark frames without any signal but must be separated from other sources of noise such as read and shot noise. Dark current is highly dependent on both exposure time and detector temperature. Furthermore, dark current in CMOS detectors is difficult to measure due to pixel to pixel variability as well as increased thermal noise for short exposure readouts \cite{Mahato2018}.

We find that the overall noise levels in the apx60 are very low, therefore we use the mean pixel value in a dark frame (i.e. no signal) as a proxy for dark current in this analysis. In a dark frame where the signal component is zero, the mean pixel value represents the build-up of time-dependent or temperature-dependent signals in the dark frame, such as dark current. Note that in most cases, the median pixel value is preferable to the mean as it is less sensitive to outlying values. However, the apx60 reports ADU in integer units and the noise levels in the camera are low enough to produce the same median pixel value across most temperature or exposure datasets. Therefore, we choose to use the mean to better probe the apx60's noise characteristics.

Using the off-the-shelf apx60 camera package, we calculate the mean pixel value for images with exposure times of 0.001, 0.01, 0.1, 1, 5, 10, 15, and 30 s. The mean pixel value is calculated across an entire, raw image. Due to the difference in ADC temperature for short exposure readouts as well as software corrections such as optical black level calibrations, bias subtraction in CMOS cameras is difficult and can result in negative dark current values. In the case of the apx60, the bias frames---i.e. the 0.001 second dark frame at a given detector temperature---tended to have a higher mean pixel value than longer duration frames, therefore we did not bias subtract when calculating dark current. The right panel of Figure \ref{fig:readnoise} shows the mean pixel values as a function of exposure time when the detector is left to rest at ambient temperature ($\sim$19 C). Thermal control was not used in this dataset. At 19 C, the mean pixel value is within the read noise for exposure $<$1 second. Above 1 second, the dark current and other time-dependent noise components are observable but maintain levels well below an electron per second and are therefore negligible in general analyses. Based on the fit to the data, time-dependent noise increases with exposure time as 0.04 electron/s.

We also attempt to measure the dark current as a function of detector temperature over the full exposure range. Using the built-in cooling electronics in the apx60, the camera can achieve a cooling $\Delta$ of -30 C. The camera does not have a heating feature, therefore ambient is the max. Dark frames were taken across the full exposure range at temperatures of -10, -5, 0, 5, 10, 15, and 20 C, where 20 C was ambient and did not use any temperature control systems. However, we find that for fast readouts ($<$15 seconds) the cooling system is unable to maintain the detector's temperature and instead the detector heats up increasing its median pixel value and dark current. This is not immediately obvious in the data as the detector is equipped with a temperature sensor which is readout in the FITS header. However, we find that the temperature readout is not regularly updated for short exposures resulting in inaccurate temperature assignments. For 0.001 second exposures taken in a continuous observing loop, we find that after reading out 20 frames, the detector stabilizes at a temperature $>$5$\degree$ the displayed temperature.

Consequently, instead of the mean pixel value linearly increasing with exposure time as would be expected, we find that the mean pixel value exponentially decreases with increasing exposure time. The degree to which the mean pixel value increases from detector heating depends on the exposure time and the set point temperature---with shorter exposure times resulting in more heating and lower temperature set points resulting in a larger difference in mean pixel values between the shortest and longest exposures.

\subsubsection{Patterned Noise}\label{subsec:patnoise}

\begin{figure}%
    \centering
    \subfloat{{\includegraphics[width=8cm]{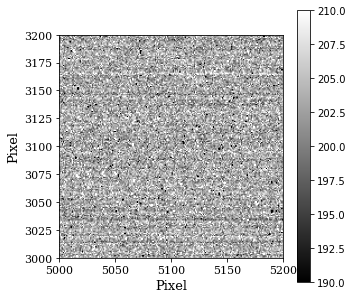} }}%
    \qquad
    \subfloat{{\includegraphics[width=7cm]{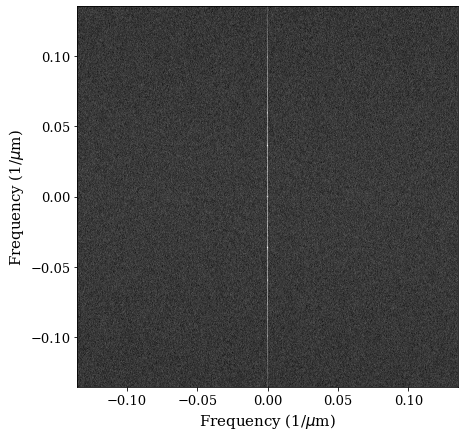} }}%
    \qquad
    \subfloat{{\includegraphics[width=13cm]{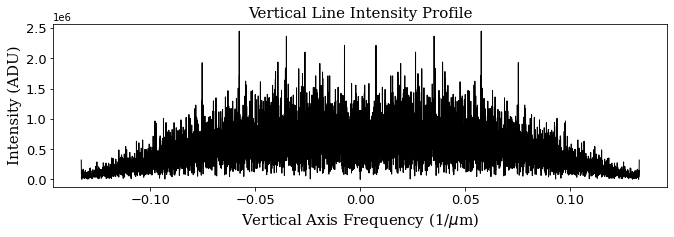} }}
    \caption{[top left] An example image from the off-the-shelf apx60 camera with an exposure time of 0.001 s, gain = 1, and detector temperature of 15C. The image is zoomed in to highlight the patterned noise. The colorbar is in units of ADU. [top right] The 2D Fourier transform of the full-frame version of the image on the left. For visualization purposes, the figure shows the log of the Fourier magnitudes. The white, vertical line at the center indicates patterned noise in the image with a horizontal striped structure. [bottom] A 1D visualization of the vertical line in the 2D Fourier transform. The intensity is given in units of ADU and represents the magnitude of the patterned noise at a given vertical spatial frequency. The spikes indicate the frequency of horizontal stripes in the raw image data.}%
    \label{fig:pattern_noise}%
\end{figure}

Images from the apx60 package exhibit patterned noise in the form of horizontal stripes which span the full image width. These bands visibly move vertically across the screen in consecutive exposures. The upper left panel of Figure \ref{fig:pattern_noise} shows the patterned noise signature in a dark frame with a 0.001 s exposure, gain of 1, and 15 C detector temperature. The horizontal lines are very fine, therefore the figure is zoomed in to x = (5000, 5200) pixels and y = (3000, 3200) pixels to emphasize the noise structure.

The upper right panel of Figure \ref{fig:pattern_noise} shows the 2D Fourier transform of the full image in the left panel. The spatial frequency is given in units of 1/$\mu$m with an input pixel size of 3.76 $\mu$m. A clear vertical line is seen at the zero frequency index. This reveals the presence of horizontal patterned noise in the raw image which takes the form of stripes. Since the line is at zero frequency (i.e. DC component) in the Fourier transform, this implies the stripes have consistent intensity in the horizontal direction. 

The bottom panel of Figure \ref{fig:pattern_noise} shows the vertical line's intensity as a function of vertical spatial frequency. The spikes in the intensity profile align with the brighter points in the vertical line shown in the upper right panel. These spikes give the spatial frequency---and subsequently the separation---of the horizontal stripes in the raw image data. The most prominent bright points occur at vertical frequencies of approximately $\pm$0.008 1/$\mu$m, $\pm$0.035 1/$\mu$m, $\pm$0.058 1/$\mu$m, and $\pm$0.075 1/$\mu$m. This corresponds to separations of 125 $\mu$m, 28 $\mu$m, 17 $\mu$m, and 13 $\mu$m. Additionally the power in the vertical line decreases at the largest spatial frequencies (smallest separations) as indicated by the dimming of line near the edges of the image. The power and structure of this line requires further investigation. 

%We calculate the power in the noise by isolating the vertical line and computing its power spectrum via:

%\begin{equation*}
%    \mathrm{PS} = \mathrm{abs}(\mathrm{DN}_{cent})^2
%\end{equation*}

%\noindent where PS is the power spectrum and $\mathrm{DN}_{line}$ is the array of values in the vertical line. The power spectrum is then summed to find the total power in the vertical line and normalized by the number of pixels in the image. We find the patterned noise in the apx60 has a power of 4.2e7 DN$^2$. This corresponds to $\sim$6.5e3 DN in patterned noise per image.

We also attempt to measure the temporal frequency of the noise using 1D Fourier transforms. However, the noise's frequency is faster than the camera's readout capabilities. Finally, the patterned noise is likely coming from the USB connection running from the camera to computer. When utilizing different USB cables, the patterned noise signature changes. To minimize the influence of patterned noise in on-orbit data, we intend to remove the USB connector on the CuRIOS-ED camera and solder data connections directly to the camera boards.

\subsection{Repackaged apx60 Thermal Vacuum Testing}\label{subsec:environ_test}

\begin{figure} [ht]
   \begin{center}
   \begin{tabular}{c} %% tabular useful for creating an array of images 
   \includegraphics[height=9cm]{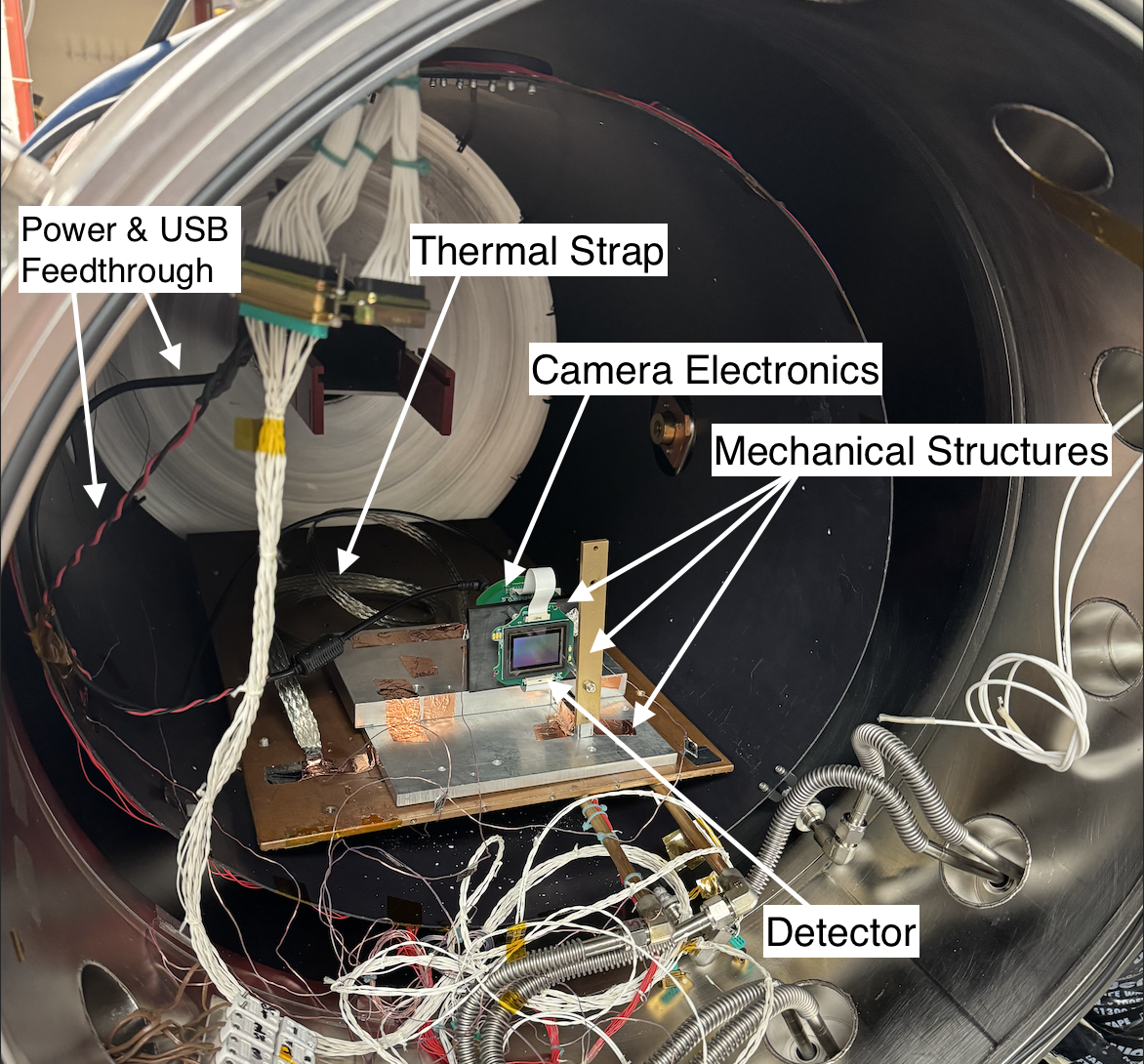}
   \end{tabular}
   \end{center}
   \caption 
   { \label{fig:TVAC_setup} 
The TVAC testing setup. The repackaged apx60 camera sits on two aluminum plates used for mechanical structure. The camera is further supported by an thin, vertical piece of alumnimum which is screwed into the aluminum plates under the camera. A thermal strap removes excess heat from an FPGA on the compute board at the back of the camera.}
\end{figure}

We performed thermal vacuum (TVAC) testing on the repackaged apx60 to prove functionality and measure performance characteristics in vacuum conditions and at a range of temperatures. A rough schematic of the repackaged apx60 can be seen in Figure \ref{fig:CAD}. The repackaged unit consists of the original Atik apx60 detector board, power board, and compute/FPGA board as well as a custom interface plate used to mount the three boards together. Figure \ref{fig:TVAC_setup} shows the thermal vacuum setup, with components labeled. The camera (electronics and interface plate) is assembled and facing the open vacuum chamber door. A thermal strap is taped to the camera's FPGA with copper tape to remove excess heat.

The temperature in the thermal vacuum chamber was controlled by heating or cooling the chamber's baseplate and shroud. Cooling was supplied through liquid nitrogen. Seven temperature sensors were used to monitor the instrument throughout testing. The temperature sensors were placed on the TVAC chamber baseplate, the TVAC chamber shroud, the camera's FPGA, the end of the FPGA thermal strap, the front of the interface plate closest to the detector, and the back of the interface plate closest to the power/FPGA boards. The two interface plate sensors are used to measure thermal gradients across the camera's interface plate. The final temperature sensor is used to readout the detector temperature and is internal to the apx60 electronics.

The camera was tested from -35 C to 40 C with an average chamber pressure of 1e-6 Torr. At the max and min temperatures, the unit was left to soak for 4 hours. The thermal cycle always began with the cold soak at -35 C and finished with the hot soak at 40 C to protect the camera electronics. We considered the camera to be at a given temperature once the detector temperature was within 2 degrees of the setpoint. The full thermal cycle was repeated three times. An additional cold soak at -35 C was also performed. Extensive data was taken on the thermal behavior/output of the system. Dark frames with exposure times of 0.001 s, 1 s, and 30 s were taken at the max and min temperatures and at 5 C intervals. The exposure mode, frame size, input voltage, and gain of these exposures was varied over the course of TVAC test to allow for performance comparisons.

The repackaged camera successfully completed TVAC testing. However, an unknown light source in the TVAC chamber polluted a large fraction of the dark frames. Therefore, characterization of the dark current and other time-dependent signals will be published in a follow-up analysis after a second TVAC run. In this paper, we present on the initial findings from these tests including read noise and the detector's thermal responses. Further analysis is ongoing.

\subsubsection{Repackaged Camera Read Noise in TVAC}\label{subsec:readnoise_environ}

We measure the read noise in the TVAC data over the full range of tested temperatures as well as for different camera exposure modes. The Atik apx60 SDK supports three modes of exposure: power save, normal, and fast. From the camera's manual, the power save mode uses only essential circuits, normal mode has all circuits powered up, and fast mode is optimized for reading out at high-speeds. At each temperature, we take 20 dark frames in each exposure mode with exposure times of 0.001 seconds and a gain of 1.

\begin{figure} [ht]
   \begin{center}
   \begin{tabular}{c} %% tabular useful for creating an array of images 
   \includegraphics[height=7cm]{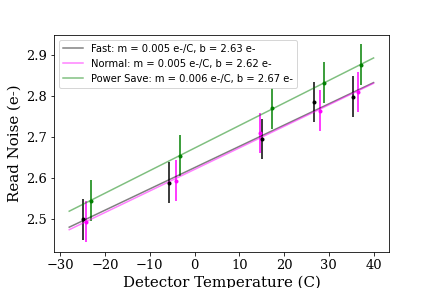}
   \end{tabular}
   \end{center}
   \caption 
   { \label{fig:readnoise_TVAC} 
Read noise as a function of detector temperature in TVAC testing. The data are taken in three exposure modes: power save, normal, and fast mode. The data are fit with a linear regression with parameters included in the legend. The sum of the SEM errors and the 0.05 e- error from the data completeness calculation (see Section \ref{subsec:readnoise}) are included for each data point. Note that the SEM errors are very small due to the large number of pixels, therefore the data completeness error dominates the error bars.}
\end{figure}

Figure \ref{fig:readnoise_TVAC} shows the read noise for each exposure mode as a function of detector temperature. The read noise is taken to be the median pixel read noise, as described in Section \ref{subsec:readnoise}. We find that power save mode has a consistently higher read noise value than fast or normal modes at all temperatures. The normal and fast mode read noise have identical values at all temperatures tested. The IMX455 tends to be read noise dominated due to its very low dark current ($<<$1 e-/s) thus we determine the normal or fast modes to be preferable for low-signal applications. 

Similar to the off-the-shelf apx60 camera characterization results, the repackaged apx60 TVAC data exhibit a weak temperature dependence in the read noise. In the case of all three exposure modes tested, the read noise value increases by $\sim$0.3 e- from -25 C to 35 C. For the same detector temperature, we find that the read noise value in the TVAC data agrees within errors with the off-the-shelf data in Section \ref{subsec:readnoise}. For example, at 0 C, the off-the-shelf apx60 data has a read noise of 2.4$\pm$0.2 e- and the TVAC data has a read noise of 2.62$\pm$0.05 e-. We note that the error on the TVAC data is smaller than the off-the-shelf data due to the larger TVAC dataset providing better statistics in the read noise calculation.

\section{Conclusions}\label{sec:conclusions}

We present the instrumental design and initial instrument testing for CuRIOS-ED---a technology demonstrator for a larger CubeSat constellation to study time-domain astrophysics. CuRIOS-ED will launch in early 2026 and function to space-qualify a low-cost commercial camera solution, validate a pointing system with $<$1" stability, and provide initial on-orbit calibration data.
We include results on the space-qualification of the Atik apx60 commercial camera package with Sony IMX455 detector. We performed baseline performance characterization on the off-the-shelf camera including measurements of baseline image intensities, read noise, dark current, and  patterned noise. We then repackaged and modified the camera to fit within a CubeSat form factor with vacuum-compatible parts. We compare the results from the thermal vacuum chamber to baseline values. The main takeaways are:

\begin{itemize}
  \item The apx60 is typically read noise dominated with a read noise value of 2.4$\pm$0.3 e- in the baseline characterization and a value of 2.7$\pm$0.2 e- in the TVAC characterization.  
  \item The apx60 exhibits patterned noise with a horizontally striped structure and spatial frequency of 6e-5 $\mu$m. Measurement of the power in this noise is ongoing.
  \item Dark current and other time-dependent noise is negligible in the apx60 at exposure times $<$1 second. At exposure times longer than 1 second, the dark current is still well below an e-/s.
  \item The camera survived four days of thermal vacuum testing with pressures below $<$5e-5 Torr (down to 5e-7 Torr) and temperatures ranging from -35 C to 40 C. The camera returned to nominal operations and detector characteristics post testing.
\end{itemize}

Further analysis of the TVAC performance is ongoing.

\acknowledgments % equivalent to \section*{ACKNOWLEDGMENTS}  

The authors would like to acknowledge Matt Dexter (UC Berkeley) and Chris Smith (SSL) for their support and guidance in executing these experiments. The authors would also like to acknowledge the StarSpec Technologies team---specifically Javier Romualdez, Jason Brown, John Hartley, and Steven Li---for their collaboration and enthusiasm for the CuRIOS project.
 
%This material is based upon work supported by the National Science Foundation Graduate Research Fellowship under Grant No. DGE 2146752. 

H.C.G. and J.R.L. acknowledge support from the National Science Foundation under grant No. DGE 2146752 and the Heising-Simons Foundation under grant No. 2022-3542.
H. C. G. acknowledges support from the H2H8 foundation.

%%%%% References %%%%%

%FOR SPIE
\printbibliography

%\bibliography{report}   % bibliography data in report.bib
%\bibliographystyle{jwapjbib}   % makes bibtex use spiejour.bst

\end{document}